\documentclass[twocolumn, prl, aps, floats]{revtex4}
\DeclareMathAlphabet{\mathpzc}{OT1}{pzc}{m}{it}

\usepackage{amsmath, amssymb, bbm, bm, color, epsf, epsfig, float, graphicx, ragged2e, relsize, setspace, tabularx, yfonts}
\def\lahigh{5ex}
\newlength{\digit}
\def\one{\hspace{\digit}}

\def\na{\overline{a}}
\def\nb{\overline{b}}
\def\nc{\overline{c}}
\def\nd{\overline{d}}
\def\Neta{\mbox{}\vspace{0pt}\hspace{0pt}		\includegraphics[height=\lahigh]{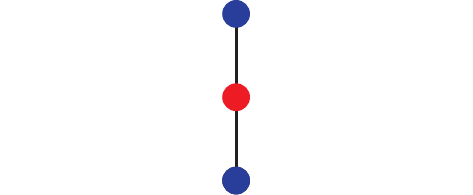}}
\def\Netaa{\mbox{}\vspace{0pt}\hspace{0pt}		\includegraphics[height=\lahigh]{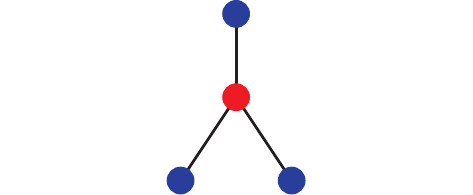}}
\def\Netaaa{\mbox{}\vspace{0pt}\hspace{0pt}		\includegraphics[height=\lahigh]{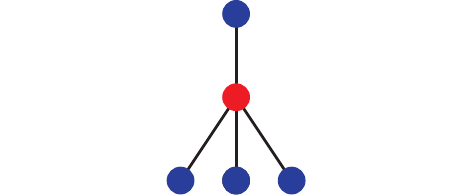}}
\def\Netab{\mbox{}\vspace{0pt}\hspace{0pt}		\includegraphics[height=\lahigh]{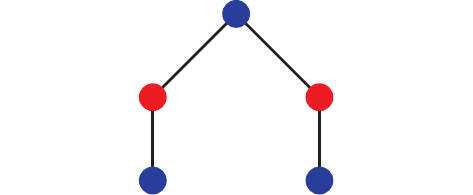}}
\def\Netabb{\mbox{}\vspace{0pt}\hspace{0pt}		\includegraphics[height=\lahigh]{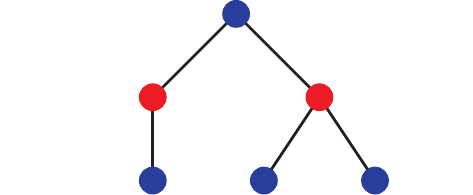}}
\def\Netaabb{\mbox{}\vspace{0pt}\hspace{0pt}		\includegraphics[height=\lahigh]{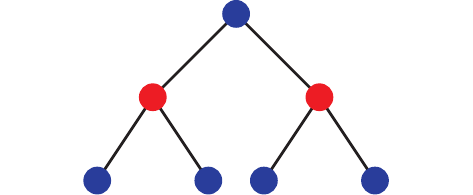}}

\def\Netabbb{\mbox{}\vspace{0pt}\hspace{0pt}		\includegraphics[height=\lahigh]{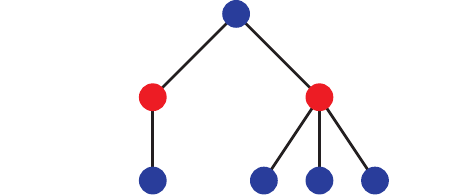}}
\def\Netaabbb{\mbox{}\vspace{0pt}\hspace{0pt}	\includegraphics[height=\lahigh]{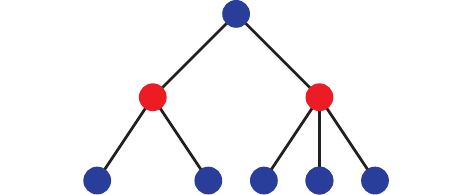}}
\def\Netaaabbb{\mbox{}\vspace{0pt}\hspace{0pt}	\includegraphics[height=\lahigh]{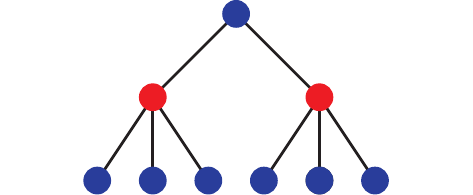}}
\def\Netabc{\mbox{}\vspace{0pt}\hspace{0pt}		\includegraphics[height=\lahigh]{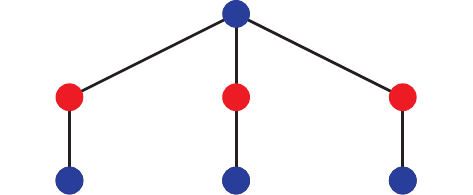}}
\def\Netabcc{\mbox{}\vspace{0pt}\hspace{0pt}		\includegraphics[height=\lahigh]{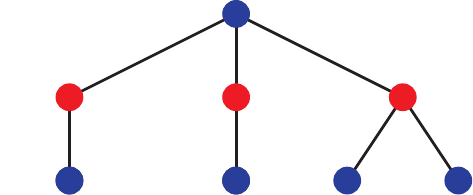}}
\def\Netabbcc{\mbox{}\vspace{0pt}\hspace{0pt}		\includegraphics[height=\lahigh]{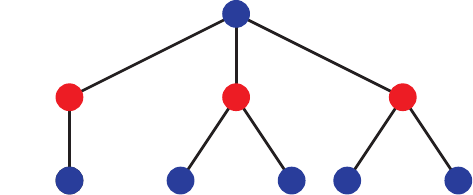}}
\def\Netabccc{\mbox{}\vspace{0pt}\hspace{0pt}		\includegraphics[height=\lahigh]{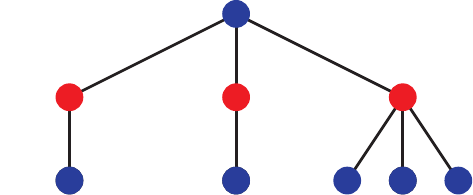}}
\def\Netabbccc{\mbox{}\vspace{0pt}\hspace{0pt}	\includegraphics[height=\lahigh]{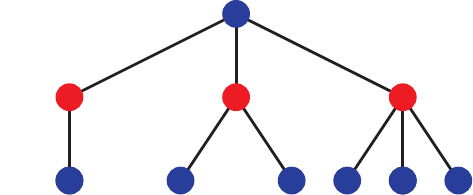}}
\def\Netabbbccc{\mbox{}\vspace{0pt}\hspace{0pt}	\includegraphics[height=\lahigh]{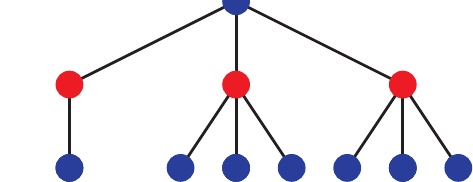}}
\def\Netaabbcc{\mbox{}\vspace{0pt}\hspace{0pt}	\includegraphics[height=\lahigh]{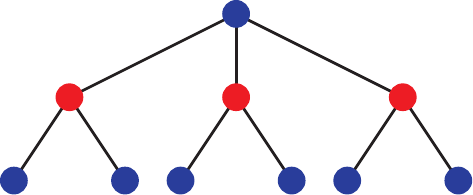}}
\def\Netaabbccc{\mbox{}\vspace{0pt}\hspace{0pt}	\includegraphics[height=\lahigh]{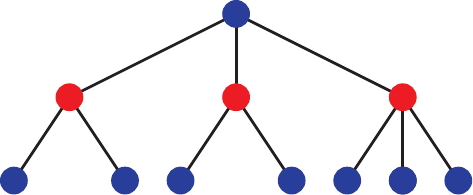}}
\def\Netaabbbccc{\mbox{}\vspace{0pt}\hspace{0pt}	\includegraphics[height=\lahigh]{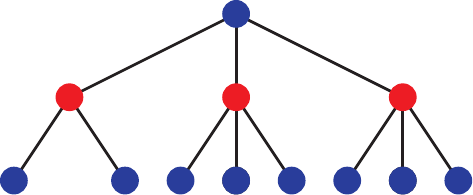}}
\def\Netaaabbbccc{\mbox{}\vspace{0pt}\hspace{0pt}	\includegraphics[height=\lahigh]{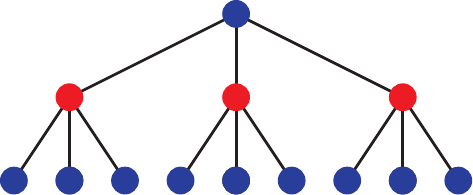}}

\def\ga{\mbox{}\vspace{0pt}\hspace{0pt}			\includegraphics[height=\lahigh]{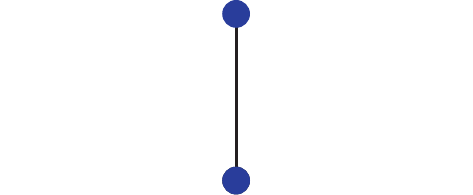}}
\def\gaa{\mbox{}\vspace{0pt}\hspace{0pt}			\includegraphics[height=\lahigh]{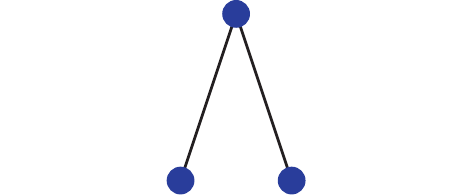}}
\def\gaaa{\mbox{}\vspace{0pt}\hspace{0pt}		\includegraphics[height=\lahigh]{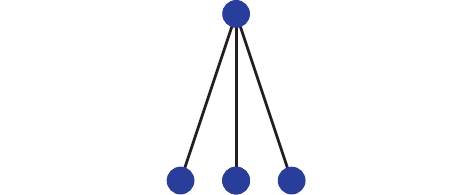}}
\def\gab{\mbox{}\vspace{0pt}\hspace{0pt}			\includegraphics[height=\lahigh]{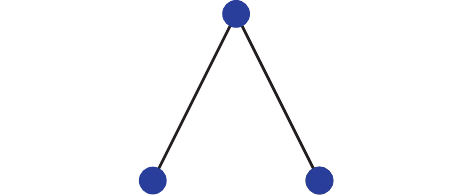}}
\def\gabb{\mbox{}\vspace{0pt}\hspace{0pt}		\includegraphics[height=\lahigh]{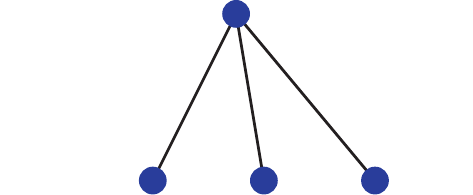}}
\def\gaabb{\mbox{}\vspace{0pt}\hspace{0pt}		\includegraphics[height=\lahigh]{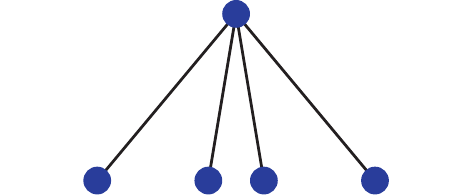}}
\def\gabbb{\mbox{}\vspace{0pt}\hspace{0pt}		\includegraphics[height=\lahigh]{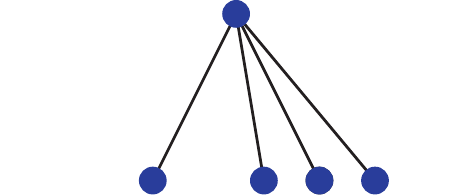}}
\def\gaabbb{\mbox{}\vspace{0pt}\hspace{0pt}		\includegraphics[height=\lahigh]{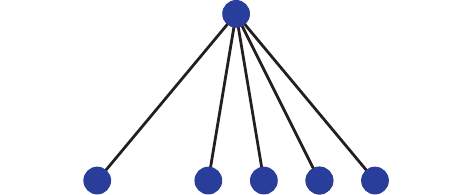}}
\def\gaaabbb{\mbox{}\vspace{0pt}\hspace{0pt}		\includegraphics[height=\lahigh]{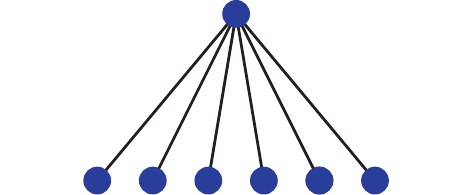}}
\def\gabc{\mbox{}\vspace{0pt}\hspace{0pt}		\includegraphics[height=\lahigh]{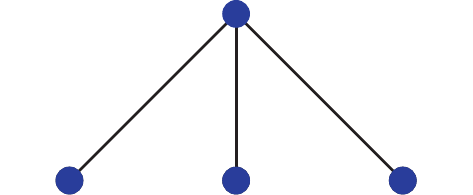}}
\def\gabcc{\mbox{}\vspace{0pt}\hspace{0pt}		\includegraphics[height=\lahigh]{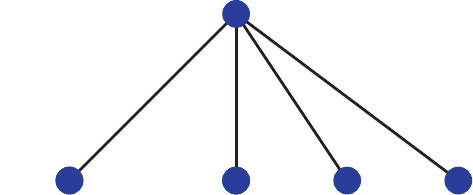}}
\def\gabbcc{\mbox{}\vspace{0pt}\hspace{0pt}		\includegraphics[height=\lahigh]{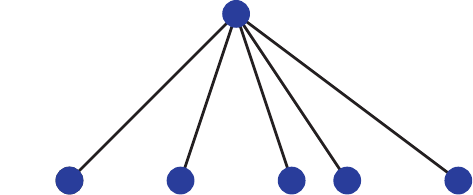}}
\def\gabccc{\mbox{}\vspace{0pt}\hspace{0pt}		\includegraphics[height=\lahigh]{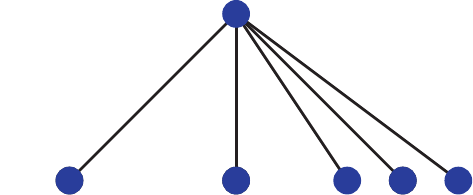}}
\def\gabbccc{\mbox{}\vspace{0pt}\hspace{0pt}		\includegraphics[height=\lahigh]{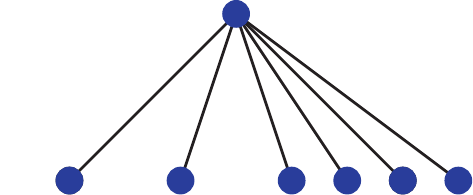}}
\def\gabbbccc{\mbox{}\vspace{0pt}\hspace{0pt}		\includegraphics[height=\lahigh]{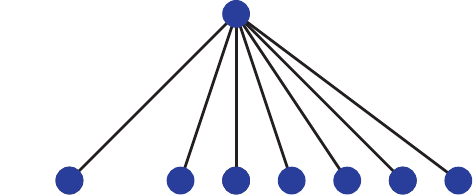}}
\def\gaabbcc{\mbox{}\vspace{0pt}\hspace{0pt}		\includegraphics[height=\lahigh]{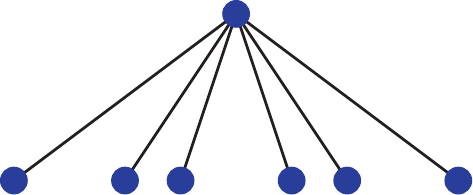}}
\def\gaabbccc{\mbox{}\vspace{0pt}\hspace{0pt}		\includegraphics[height=\lahigh]{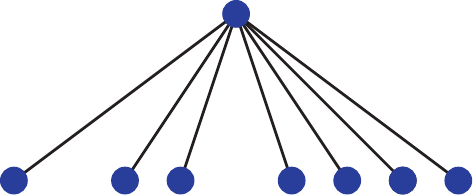}}
\def\gaabbbccc{\mbox{}\vspace{0pt}\hspace{0pt}	\includegraphics[height=\lahigh]{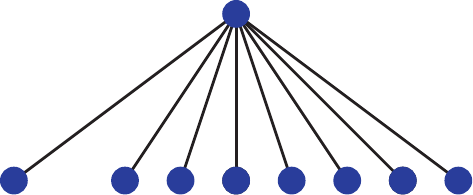}}
\def\gaaabbbccc{\mbox{}\vspace{0pt}\hspace{0pt}	\includegraphics[height=\lahigh]{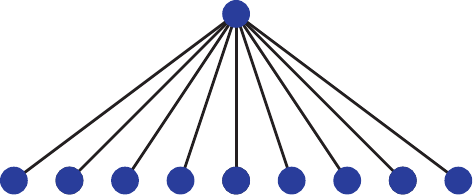}}

\begin{document}
\begin{small}

		\title{\flushleft
			{\sf\LARGE \textbf{Biological logics are restricted}} \\
			\vspace{5pt}
			{\sf \textbf{T. M. A. Fink and R. Hannam}} \\
			\vspace{3pt} 	
			{\sf \small \textbf{London Institute for Mathematical Sciences, Royal Institution, 21 Albermarle St, London W1S 4BS, UK}} \\
			\mbox{}				\vspace{-20pt} 
			\mbox{}
		}
		\maketitle
\noindent
\begin{justify}
{\sf\textbf{Networks of gene regulation govern morphogenesis, determine cell identity and regulate cell function. 
But we have little understanding, at the local level, of which logics are biologically preferred or even permitted. 
To solve this puzzle, we studied the consequences of a fundamental aspect of gene regulatory networks: genes and transcription factors talk to each other but not themselves. 
Remarkably, this bipartite structure severely restricts the number of logical dependencies that a gene can have on other genes.
We developed a theory for the number of permitted logics for different regulatory building blocks of genes and transcription factors. 
We tested our predictions against a simulation of the 19 simplest building blocks, and found complete agreement. 
The restricted range of biological logics is a key insight into how information is processed at the genetic level.
It constraints global network function and makes it easier to reverse engineer regulatory networks from observed behavior.
}}
\end{justify}
\mbox{} \\ \noindent
{\sf\textbf{\textcolor{red}{\large Introduction}}} \\
The development and maintenance of living organisms requires a lot of computation. 
This is mainly performed at the molecular level through gene regulatory networks.
They govern the creation of body structures, regulate cell function, and are responsible for the progression of diseases.
Since the landmark discovery of induced pluripotent stem cells, scientists have identified special combinations of transcription factors which control cell identity \cite{Pawlowski2017, Kamao2014}.
Precision control over cell fate opens up the possibility of manufacturing cells for 
drug development \cite{Engle2013},
disease modelling \cite{Kanherkar2014a} and
regenerative medicine \cite{Cherry2013}. 
\\ \indent 
Models of gene regulatory networks have been investigated for half a century \cite{Kauffman1969,Huang2005}. 
They actually predate aspects of our understanding of gene regulation itself, such as the role of transcription factors.
Partly because of this, models of regulatory networks have tended to be overly simplistic \cite{Bornholdt2008}---disregarding, for example,
the complementary roles played by genes and transcription factors \cite{Hannam2019,Hannam2019b}.
Despite the simplicity of these models, however, a theoretical understanding of their typical behavior proved elusive until the mid-2000s \cite{Socolar2003,Samuelsson2003,Shmulevich2004,Mihaljev2006}. 
\\ \indent 
During the 20th century, various problems in biology have transitioned from a descriptive \cite{Reed2004} to a predictive science \cite{Reed2015}.
Examples include protein folding \cite{Jumper2021,Ahnert2015}, genotype-phenotype maps \cite{Wagner2008} and the segmentation of vertebrates \cite{Gomez2008}.
Yet a predictive understanding of genetic computation remains elusive, despite intense interest from researchers across fields \cite{Rand2021,Yan2017}.
\\ \indent 
Predictability comes from mathematical structure, and mathematical structure is the consequence of constraints.
In physics, there is an abundance of constraints, typically expressed in the form of conservation laws. 
The role of constraints in biology is less well understood, but modularity \cite{Wagner2007,Ahnert2016} and symmetry \cite{,Dingle2018,Johnston2022} seem to play important roles.
\\ \indent One constraint on gene regulatory networks that is hiding in plain site is their bipartite nature: genes and transcription factors talk to each other but not themselves.
A transcription factor is a protein or complex of proteins which are themselves synthesized from expressed genes.
Thus a transcription factor depends on one or more genes.
A gene is a particular segment of DNA that codes for a protein, flanked by one or more binding sites for different transcription factors. 
These transcription factors promote or block the transcription of the gene.
Thus a gene depends on one or more transcription factors. 
In this way, the expression levels of genes are determined by those of other genes, but only indirectly, via transcription factors \cite{Buccitelli2020}.
Bipartite models of regulation can reflect key biological details, such as different gene and transcription factor connectivities \cite{Hannam2019,Hannam2019b}.
\\ \indent
Our familiarity with the bipartite constraint belies its importance in determining function.
As we shall see, it severely restricts the range of logical dependencies that any one gene can have on other genes.
Identical arguments apply to the dependencies that a transcription factor can have on other transcription factors, but for brevity we stick to genes.
\\ \indent In this article we do four things.
First, we enumerate the different regulatory motifs that relate one gene to other genes via transcription factor middlemen. 
Gene regulatory networks are built out of these regulatory motifs, the 19 simplest of which are shown in Fig. \ref{TableCompositions}.
Second, we develop an exact theory for the number of permitted gene-gene logics, for any regulatory motif.
This number tends to be much smaller than the number of possible gene-gene logics.
Third, for the 19 simplest regulatory motifs, we compared our prediction to a simulation of how  gene and transcription factor logics combine, and found exact agreement.
Fourth, we quantify the benefits of this restriction on logics for reverse engineering gene regulatory circuits from experiments.
\\ \\ \noindent {\sf\textbf{\textcolor{red}{\large Results}}} 
\\ \noindent {\sf\textbf{\textcolor{black}{A puzzle}}} 
\\ Our key insight is that the bipartite nature of gene regulatory networks 
severely limits the number of logical dependencies that one gene can have on other genes.
To understand this conceptually, consider a social network puzzle.
Imagine that men and women talk to the opposite sex but not their own, and each person has one of only two moods, happy or sad.
As a man, your own mood depends on the moods of two women.
For instance, you might be happy only if both women are happy.
Or you might copy the mood of the first and ignore the second.
The mood of each woman depends, in turn, on the mood of two men.
So, ultimately, your mood is governed by the mood of the four men.
In how many ways can your mood depend on them?
\\ \indent 
You might guess that there are $2^{2^4}$ = 65,536 ways, which is the number of logical dependencies on four variables.
But in reality there are just 520 ways to depend on the four men.
The hidden variables of the women greatly reduces the range of logical dependencies. 
\\ \indent 
The solution to this puzzle hints at a fundamental aspect of dynamical systems in which two species depend on each other but not themselves. 
It suggests that the logical dependencies observed between a single species are highly restricted. 
The preeminent example of such a system is gene regulatory networks, in which genes interact via transcription factors.
As we shall see, the number of permissible gene-gene logics is greatly reduced.
\\ \\ {\sf\textbf{\textcolor{black}{Theory}}} \\
Before we calculate the number of permissible gene-gene logics---which we call biological logics---we review some general properties of logics.
Logics are also known as Boolean functions, and we use the terms interchangeably.
There are $2^{2^n}$ logics of $n$ variables.
For $n=2$, they are
true, false, 
$a$, $b$, $\na$, $\nb$, 
$ab$, $a\nb$, $\na b$, $\na\nb$,
$a+b$, $a+\nb$, $\na+b$, $\na+\nb$,
$ab+\na\nb$ and $a\nb+\na b$.
In this notation, $\na$ means {\sc not} a, $ab$ means $a$ {\sc and} $b$, and $a+b$ means $a$ {\sc or} $b$.
\\ \indent
Notice that two of these functions depend on no variables (true and false), 
four depend on one variable ($a$, $b$, $\na$ and $\nb$), 
and the rest depend on two variables. 
Let $s(n)$ be the number of logics of $n$ variables that depend on all $n$ variables.
By the principle of inclusion and exclusion, 
\begin{equation}
    s(n) = \sum_{i=0}^n (-1)^{n-i} \textstyle \binom{n}{i}2^{2^i}.
    \label{AllVariableLogics}
\end{equation}	
The first few $s(n)$ are 2, 2, 10, 218, 64594, starting at $n=0$ (OEIS A000371 \cite{Sloane}).
\def\vs{\vspace{0.014in}}
\def\lahigh{4.2ex}
\begin{figure}[b!]
\includegraphics[width=0.9\columnwidth]{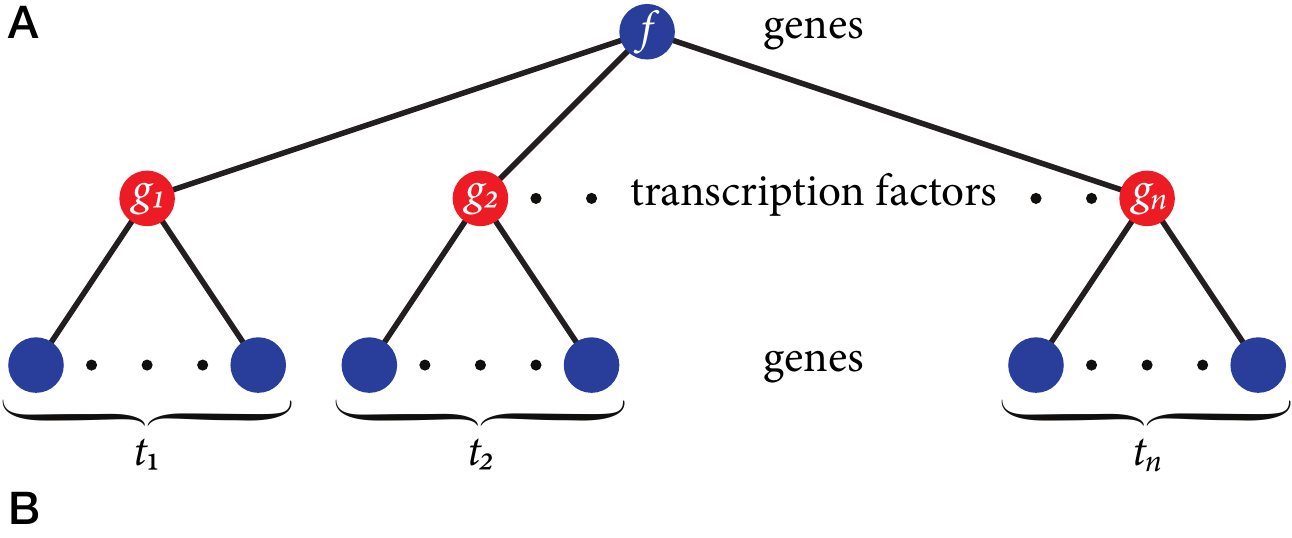}
\setlength{\tabcolsep}{0pt}
\begin{tabularx}{\columnwidth}{@{\extracolsep{\fill}}cccccc}
\emph{}		&					& \emph{Biological} 	&& \emph{Projected} 	& \emph{All}					\\	
\emph{Short-}	& \emph{Regulatory} 	& \emph{logics} 	&& \emph{regulatory} 	& \emph{logics}					\\
\emph{hand}	& \emph{motifs} 		& $c(t_1,\ldots,t_n)$ 	&& \emph{motifs}		& $2^{2^{t_1+ \ldots + t_n}}$		\\									
 $\{1\}$		& \Neta				&  4				&& \ga  	 			& 4        					\vs	\\
$\{2\}$		& \Netaa				& 16   			&& \gaa				& 16        					\vs	\\
$\{3\}$ 		& \Netaaa				& 256   			&& \gaaa				& 256        				\vs	\\
$\{1,1\}$ 		& \Netab				& 16    			&& \gab				& $16$   		   			\vs	\\
$\{1,2\}$		& \Netabb				& 88     			&& \gabb				& $256$   					\vs	\\
$\{2,2\}$		&\Netaabb			& 520    			&& \gaabb				& $65,536$   	 			\vs	\\
$\{1,3\}$		& \Netabbb			& 1528   	  		&& \gabbb				& $65,536$                       		\vs	\\
$\{2,3\}$		& \Netaabbb			& 9160 			&& \gaabbb			& $4.3 \times 10^9$ 			\vs	\\
$\{3,3\}$		& \Netaaabbb			& 161,800    	  	&& \gaaabbb			& $1.8 \times 10^{19}$ 		\vs	\\
$\{1,1,1\}$		& \Netabc				& 256 		 	&& \gabc				& 256         	  			\vs	\\
$\{1,1,2\}$		& \Netabcc			& 1696   			&& \gabcc				& 65,536      			  	\vs	\\
$\{1,2,2\}$		& \Netabbcc			& 11,344   		&& \gabbcc			& $4.3 \times 10^9$   	 	\vs	\\
$\{1,1,3\}$ 	& \Netabccc			& 30,496  			&& \gabccc			& $4.3 \times 10^9$       		\vs	\\
$\{2,2,2\}$		& \Netaabbcc			& 76,288 			&& \gaabbcc 			& $1.8 \times 10^{19}$  		\vs	\\
$\{1,2,3\}$		& \Netabbccc			& 204,304 		&& \gabbccc			& $1.8 \times 10^{19}$		\vs	\\
$\{2,2,3\}$ 	& \Netaabbccc			& 1,375,168		&& \gaabbccc			& $3.4 \times 10^{38}$ 	 	\vs 	\\
$\{1,3,3\}$ 	& \Netabbbccc			& 3,680,464 		&& \gabbbccc			& $1.8 \times 10^{19}$       	\vs	\\
$\{2,3,3\}$ 	& \Netaabbbccc		& 24,792,448 		&& \gaabbbccc			& $1.2 \times 10^{77}$  	 	\vs	\\
$\{3,3,3\}$ 	& \Netaaabbbccc    		& 447,032,128		&& \gaaabbbccc		& $1.3 \times 10^{154}$			
\end{tabularx}%
\caption{\footnotesize
\textbf{The number of gene-gene logical dependencies for the 19 simplest regulatory motifs and their projections.}
\textbf{A}
In a regulatory motif, a gene (blue) depends on $n$ transcription factors (red), which depend on $t_1,t_2,\ldots,t_n$ genes (blue), respectively. 
As a shorthand for this structure, we write $\{t_1,\ldots,t_n\}$, which counts the number of genes in the $n$ branches of the tree.
\textbf{B}
For the 19 simplest regulatory motifs, we show the number of logical dependencies $c(t_1,\ldots,t_n)$ (left) that a gene can have on other genes.
This tends to be much smaller than the number of logical dependencies in the absence of the transcription factor middlemen, $2^{2^{t_1+ \ldots + t_n}}$ (right), when the regulatory motif is projected.
}
\label{TableCompositions}
\end{figure}
\\ \indent The biological equivalent of our social network puzzle is a gene that depends on two transcription factors, each of which depends on two genes.
This is the sixth regulatory building block, or regulatory motif, in Fig. \ref{TableCompositions}B.
A regulatory motif is the connectivity that a gene has with other genes via transcription factor middlemen.
For a gene that depends on $n$ transcription factors, each of which depends on $t_1,\ldots,t_n$ genes (Fig. \ref{TableCompositions}A),  
we use as a shorthand for the regulatory motifs $\{t_1,\ldots,t_n\}$, which counts the number of genes in the $n$ branches of the tree.
\begin{figure*}[t!]
\begin{tabularx}{\textwidth}{@{\extracolsep{\fill}}llll}
\multicolumn{2}{l}{\textbf{Biological logics for \!\!\!\!\!\! \Netaabb}}																										\vspace{0.04in}			\\ 
\emph{0 variables}	 				& \emph{3 variables}							& \emph{4 variables}						& \emph{4 variables}					 								\\ 
1 $\cdot$ 	 true		 				& 8 $\cdot$ $a b c$							&16 $\cdot$ $a b c d$					& \one 8 $\cdot$ $a b (c d + \nc \nd)$									\\
1 $\cdot$ 	 false					& 8 $\cdot$ $a b + c$						&16 $\cdot$ $a b + c d$					& \one 8 $\cdot$ $(a b + \na \nb) c d$									\\
								& 8 $\cdot$ $(a+b) c$						&16 $\cdot$ $a b (c+ d)$					& \one 8 $\cdot$ $a b + c d + \nc \nd$								 	\\
\emph{1 variable}					& 8 $\cdot$ $a + b + c$						&16 $\cdot$ $(a+b) c d $					& \one 8 $\cdot$ $a b + \na \nb + c d$				 					\\
2 $\cdot$ $a$						& 8 $\cdot$ $(a+b) c + \na \nb \nc$				&16 $\cdot$ $a + b + c d$					& \one 8 $\cdot$ $a + b + c d + \nc \nd$ 	 								\\
								& 4 $\cdot$ $(a b + \na \nb) c$					&16 $\cdot$ $a b + c + d$					& \one 8 $\cdot$ $a b + \na \nb + c + d$	 								\\
\emph{2 variables}					& 4 $\cdot$ $a b + \na \nb + c$	 				&16 $\cdot$ $a + b + c + d$				& \one 8 $\cdot$ $(a b + \na \nb) (c + d)$ 	 								\\
4 $\cdot$ $a b$						& 2 $\cdot$ $(a b + \na \nb) c + (a \nb+ \na b) \nc$	&16 $\cdot$ $(a + b)(c + d)$				& \one 8 $\cdot$ $(a + b) (c d + \nc \nd)$ 	 								\\
4 $\cdot$ $a+b$					& 										&16 $\cdot$ $a b (c + d) + (\na + \nb) \nc \nd$	& \one 8 $\cdot$ $(a b + \na \nb) (c + d) + (a \nb + \na b) \nc \nd $  				\\
2 $\cdot$ $a b + \na \nb$				&										&16 $\cdot$ $(a+b)(c+d) + \na\nb\nc\nd$		& \one 8 $\cdot$ $(a + b) (c d + \nc \nd) + \na \nb (c \nd + \nc d)$ 				\\
 								&										& \one 4 $\cdot$ $a b + \na \nb + c d + \nc \nd$	& \one 2 $\cdot$  	$(a b + \na \nb) (c d + \nc \nd) + (a \nb + \na b) (c \nd + \nc d)$	\\
								&										& \one 4 $\cdot$ $(a b + \na \nb) (c d + \nc \nd )$					 				
\end{tabularx}
\caption{\footnotesize
\textbf{Biologically valid gene-gene logics for a gene which depends on two transcription factors, each of which depends on two genes.}
Of the $2^{2^4}$ = 65,536 logics of four variables, only 520 are biologically valid. 
In our notation, $\na$ means {\sc not} a, $ab$ means $a$ {\sc and} $b$, and $a+b$ means $a$ {\sc or} $b$.
We group together logics that depend on $m=$ 0, 1, 2, 3 and 4 variables. 
We need not show all 520 logics because of two kinds of symmetry.
In the first, there are $\binom{4}{m}$ ways to choose the $m$ variables.
For example, for $m=2$, there are 6 choices of two variables: $a \, b$; $a \, c$; $a \, d$; $b \, c$; $b \, d$; and $c \, d$. 
But the structure of the logics is the same for all choices, and we only show the logics for $a \, b$.
The second symmetry is given by the number before each logic.
It is the number of logics when none or some of its variables are everywhere replaced by their negation, since doing so does not change the structure of the logic.
For example, applying this to $a b$ gives $a b, a\nb, \na b$ and $\na\nb$, so we put a 4 in front of the logic $ab$, and don't show the rest.
The column sums are 2, 2, 10, 50 and 250, and $2 \binom{4}{0} + 2 \binom{4}{1} + 10 \binom{4}{2} + 50 \binom{4}{3} + 250 \binom{4}{4} = c(2,2) = 520$.
\label{Valid}
}
\end{figure*}
\begin{figure}[b!]\setlength{\hfuzz}{1.1\columnwidth}
	\begin{minipage}{\textwidth}
	\includegraphics[width=1\textwidth]{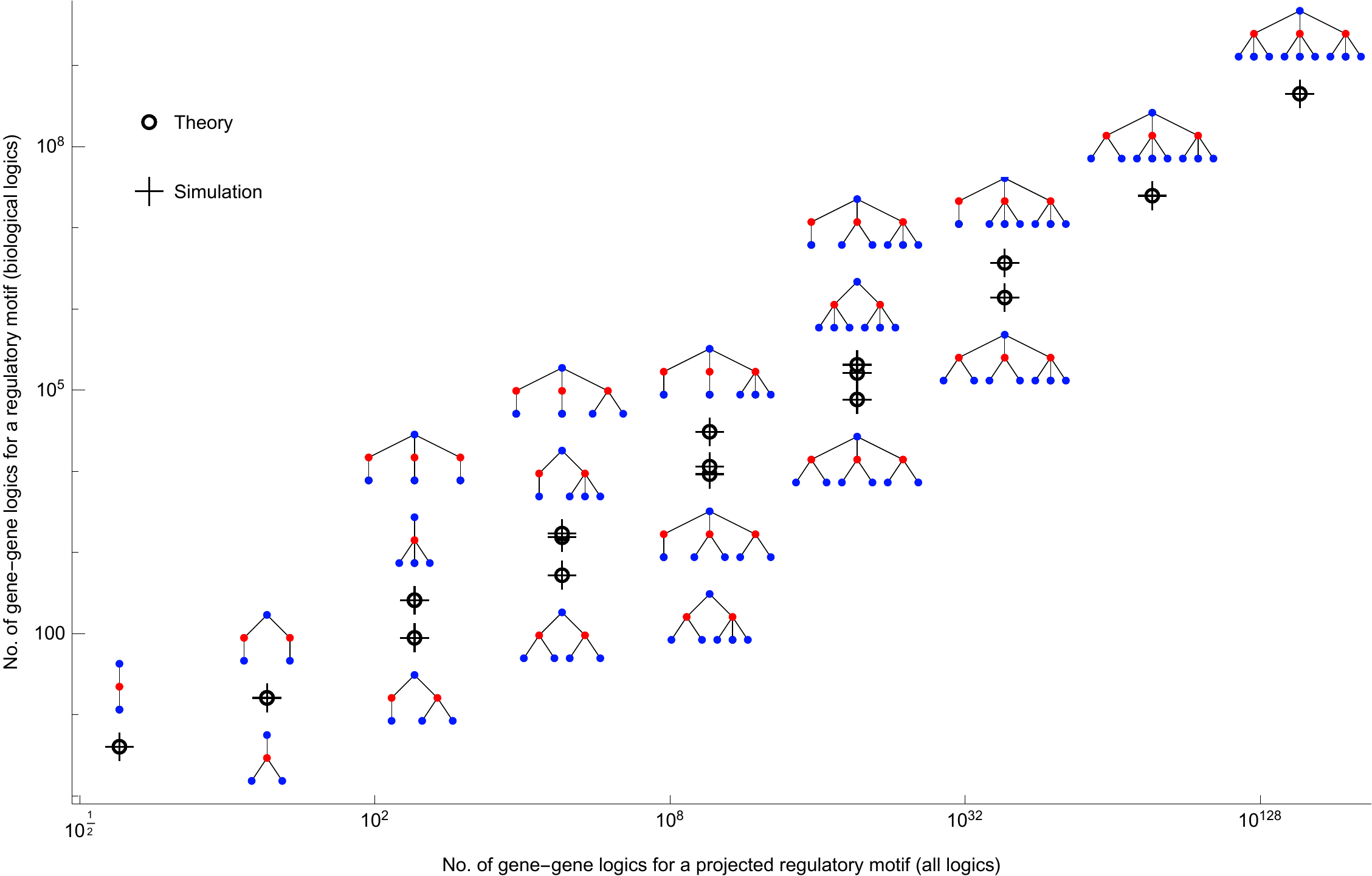}
	\caption{\footnotesize
	\textbf{Theory and simulation for the number of logics for the 19 simplest regulatory motifs.}
	Our theoretical prediction (circles) is perfectly confirmed by computer simulation (crosshairs).
	Here we have plotted each regulatory motif according to the number of gene-gene logics allowed by it and its projection---in other words, in the presence and absence of the transcription factor middlemen.
	These are the numbers on the left and right of Fig. \ref{TableCompositions}B.
	}
	\label{FigCompare}
	\end{minipage}
\end{figure}
\\  \indent 
Our main quantitative result, which we derive in the Methods, is an expression for the exact number of biological (permissable gene-gene) logics for any regulatory motif $c(t_1,\ldots,t_n)$.
(For convenience, we drop the braces around $\{t_1,\ldots,t_n\}$ when it is the argument of a function.)
The result is
\begin{equation}
    c(t_1,\ldots,t_n)  =  \sum_{m=0}^n s(m)	\!\! \sum_{\sigma_1\ldots \sigma_m}   \, \alpha_{\sigma_1} \ldots \alpha_{\sigma_m},     
    \label{CA}
\end{equation}
where
\begin{equation*}
\alpha_i =  (2^{2^i}-2)/2.
\end{equation*}
The second sum in eq. (\ref{CA}) is over all of the subsets of size $m$ of the set $\{t_1, \ldots, t_n \}$. 
Eq. (\ref{CA}) is simpler than it looks, as some examples illustrate:
\begin{eqnarray*}
    c(i)        	&=& 2 + 2 \, \alpha_i, \\
    c(i,j)     		&=& 2 + 2(\alpha_i + \alpha_j)              + 10 \, \alpha_i \alpha_j, \\
    c(i,j,k)   		&=& 2 + 2(\alpha_i + \alpha_j + \alpha_k)   \\
    			&+& 10 \big(\alpha_i \alpha_j + \alpha_j \alpha_k + \alpha_i \alpha_k \big) + 218 \, \alpha_i \alpha_j \alpha_k.
\end{eqnarray*}
With this, it is easy to calculate the number of biological logics for all of the regulatory motifs in Fig. \ref{TableCompositions}.
Let's calculate $c(2,2)$.
Since $\alpha_2 = (2^{2^2} - 2)/2 = 7$, $c(2,2) = 2 + 2(7+7) + 10 \cdot 7^2 = 520$.
These 520 biological logics are given explicitly in Fig. \ref{Valid}.
\\ \indent
Fig. \ref{TableCompositions}B shows the number of gene-gene logics for the 19 simplest regulatory motifs (left) and their projections (right).
A projection is the connectivity that results from dropping the transcription factors and connecting the genes directly to each other.
The numbers on the left tend to be vastly smaller than those on the right.
The presence of the transcription factor middlemen severely restricts the number of permissible gene-gene logics.
\\ \\ {\sf\textbf{\textcolor{black}{Simulation}}} \\
To test our prediction for the number of biological logics in eq.\ (\ref{CA}), 
we wrote a computer simulation to compute how different gene and transcription factor logics combine.
For a given regulatory motif, we assigned all possible logics to the gene at the top and to the transcription factor middlemen---$f$ and the $g_i$ in Fig. \ref{TableCompositions}A.
We call different combinations of logics equivalent if they produce the same logical dependence of the top gene on the bottom genes. 
We simulated the architectures in Fig. \ref{TableCompositions}B and compared them to our prediction, and found exact agreement, as shown in Fig. \ref{FigCompare}.
\\ \indent
In general, only a tiny fraction of all logics are biological logics, for a given regulatory motif. 
To gain some sense for which logics belong to this select group, we show them explicitly for the regulatory motif $\{2,2\}$ in Fig. \ref{Valid}.
Of the possible $2^{2^4}$ = 65,536 logics of four variables, only 520 are biological. 
\newpage \mbox{} \newpage \mbox{} \newpage 
\noindent {\sf\textbf{\textcolor{red}{\large Discussion}}}
\\ The restriction of biological logics can be understood as the consequence of two things.
First, most logics cannot be written as a composition of logics, where the composition structure reflects the regulatory motif.
Second, the assignment of logics to genes and transcription factors is redundant, in the sense that different combinations produce the same gene-gene dependence.
We consider each of these in turn before discussing the implications for reverse engineering regulatory networks.
\\  \\ {\sf\textbf{Restriction of gene-gene logics}} \\
Although we have shown that, for a given regulatory motif, most gene-gene logics are not permitted, is not self-evident which ones make the cut.
For example, for the motif $\{2,2\}$ (Fig. \ref{Valid}),
the logic $ab+cd$ is permitted, meaning that the dependent gene is expressed if $a$ {\sc and} $b$ are expressed, {\sc or}  $c$ {\sc and} $d$ are expressed.
But $ac+bd$ is not permitted---swapping $a$ and $c$ in the valid logic invalidates it.
\\ \indent
Ultimately, the condition for a valid logic is being able to express it as a composition of logics.
For the 19 simplest regulatory motifs, brute force enumeration is sufficient to determine the biologically permitted logics.
However, there are some shortcuts for going about this for these and more complex regulatory motifs.
For example, one condition for a logic to be valid for $\{2,2\}$ is that swapping the genes in either branch does not change the logic,
as is the case for $ab + cd$.
But this is not sufficient: $(a+b)c$ is permitted, but $(a+b)c + ab$ is not, even though swapping $a$ and $b$ leaves both unchanged.
Further investigation will likely uncover more comprehensive tests for biological logics.
\begin{figure}[b!]
	\centering
	\includegraphics[width=0.93\columnwidth]{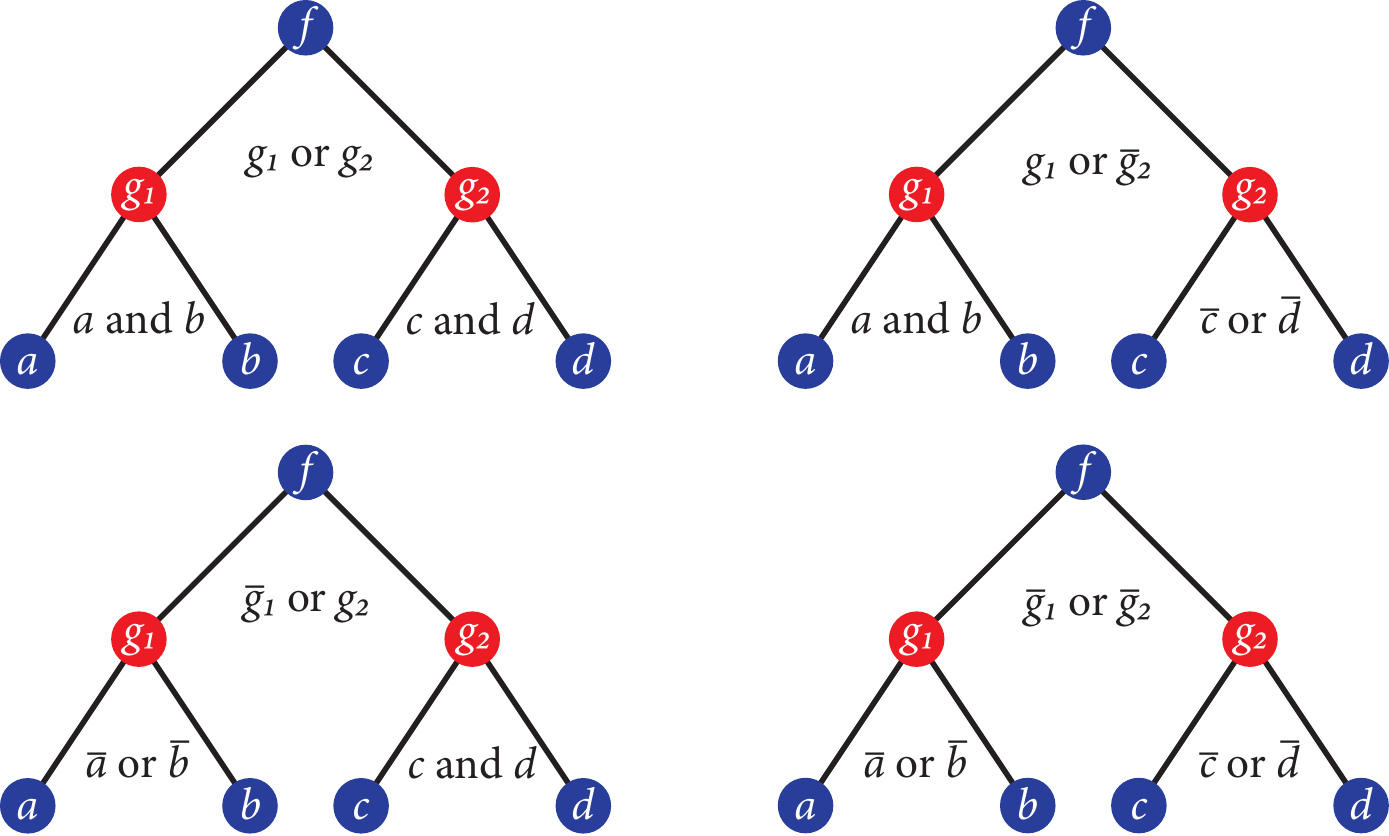}
	\caption{\footnotesize
	\textbf{Different assignments of logics to genes and transcription factors compose to give the same dependence of a gene on other genes.}
	For a gene which depends on two transcription factors, each of which depends on two genes, 
	there are four ways to achieve the logic $f$ = ($a$ {\sc and} $b$) {\sc or} ($c$ {\sc and} $d$).
	}
    \label{Degenerate}
\end{figure}
\def\vs{\vspace{0.025in}}
\def\lahigh{3.7ex}
\begin{figure}[b!]
\setlength{\tabcolsep}{0pt}
\begin{tabularx}{\columnwidth}{@{\extracolsep{\fill}}cccccc}
\multicolumn{6}{c}{\textbf{Information in bits required to reverse engineer} }				\\
\multicolumn{3}{c}{\emph{Regulatory motifs} ($I_{\rm r}$)}		& \multicolumn{3}{c}{\emph{Proj. regulatory motifs} ($I_{\rm p}$)}	\\
					& \emph{10\! genes}	&  \emph{100\! genes}	& 		& \emph{10\! genes}	&  \emph{100\! genes} \vs	 \\
\Neta				& 5				& 9					&\ga			& 5					& 9 		\vs \\	
\Netaa				& 9				& 16					&\gaa		& 9					& 16		\vs \\	
\Netaaa				& 15				& 25					&\gaaa		& 15					& 25		\vs \\	
\Netab				& 10				& 17					&\gab		& 9					& 16		\vs \\	
\Netabb				& 15				& 25					&\gabb		& 15					& 25		\vs \\	
\Netaabb				& 19				& 34					&\gaabb		& 24					& 38		\vs \\	
\Netabbb				& 20				& 34					&\gabbb		& 24					& 38		\vs \\	
\Netaabbb				& 24 				& 43 					&\gaabbb		& 40					& 58		\vs \\	
\Netaaabbb			& 29				& 52 					&\gaaabbb	& 72					& 94		\vs \\	
\Netabc				& 17 				& 28 					&\gabc		& 15 					& 25		\vs \\	
\Netabcc				& 22				& 36					&\gabcc		& 24					& 38		\vs \\	
\Netabbcc				& 26				& 45					&\gabbcc		& 40					& 58		\vs \\	
\Netabccc				& 27				& 45					&\gabccc		& 40					& 58		\vs \\	
\Netaabbcc			& 30	 			& 53	 				&\gaabbcc	& 72 					& 94		\vs \\	
\Netabbccc			& 31	 			& 54	 				&\gabbccc		& 72 					& 94		\vs \\	
\Netaabbccc			& 35 				& 62	 				&\gaabbccc	& 135 				& 162	\vs \\	
\Netabbbccc			& 36 				& 63	 				&\gabbbccc	& 135 				& 162	\vs \\	
\Netaabbbccc			& 39 				& 71	 				&\gaabbbccc	& 261 				& 293	\vs \\	
\Netaaabbbccc 			& 43 				& 80	 				&\gaaabbbccc	& 515 				& 553	\vs 	
\end{tabularx}%
\caption{\footnotesize
\textbf{The information in bits required to reverse engineer regulatory motifs and their projections.}
To reverse-engineer a regulatory motif or its projection, we need to deduce two things: the connectivity and the logic.
The number of possible connectivities depends on the size of the entire regulatory network, so we give examples for networks of 10 and 100 genes.
In general, the bipartite structure of regulatory motifs means that significantly less information is required to reverse engineer them than their projections.
}
\label{Info}
\end{figure}
\\  \\ {\sf\textbf{Redundancy of gene-gene logics}} \\
We know that the number of biological logics can be at most the number of assignments of logics to $f$ and the $g_i$ in Fig. \ref{TableCompositions}A.
But we observe that the number of biological logics is less than this.
This is because different assignments of logics to genes and transcription factors can compose to give the same gene-gene logic.
For example, for the regulatory motif $\{2,2\}$ in Fig. \ref{Valid}, 4,096 assignments compose to 520 logics.
An example of different assignments of logics which compose to the same logic is given in Fig. \ref{Degenerate}.
\\ \indent The composition of logics is a preeminent testbed for understanding input-output maps.
Many input-output maps in nature and mathematics are many-to-one, but with a non-uniform redundancy that is exponentially biased towards simple outputs \cite{Dingle2018}.
Examples include RNA secondary structure, protein complexes and model gene regulatory networks \cite{Johnston2022}.
Preliminary evidence suggests that composed logics are similarly biased towards simple logics.
Developed further, our theory for the composition of logics presents an opportunity to give a mathematical explanation of this widely observed empirical trend.
\\ \indent Looking farther afield, while we studied the composition of logics over only two levels, we believe it is possible to generalize our results to multiple levels. 
This could give theoretical backing to computational insights into the robustness and evolvability of logic gates, deftly studied by Andreas Wagner and his co-author in the context of genotype-phenotype maps \cite{Raman2011}.
When the number of composition levels is large, the limiting distribution of logics could shed light on the space of functions in some types of neural networks \cite{Mozeika2020}.
\\ \\ \noindent {\sf\textbf{Reverse engineering gene regulation}} 
\\ The reduction in the number of gene-gene logics restricts the range of global behavior of gene regulatory networks. 
A bound on the range of global behavior is the amount of information required to reverse engineer the regulatory network that gives rise to it.
Reverse engineering is a major goal of systems biology \cite{Yan2017}, and advances in methods for doing so are highly sought.
\\ \indent
As we noted earlier, a global network can be broken down piecewise into its constituent regulatory motifs.
The information required to reverse engineer the whole is the sum of the information required to reverse engineer the parts.
Let's work out the information required to reverse engineer a regulatory motif, on the one hand, and a projected regulatory motif, on the other (Fig. \ref{Info}).
On the face of it, regulatory motifs are more intricate, and ostensibly harder to reverse engineer.
But, as we shall see, the opposite is true.
 \\ \indent
The information required to reverse engineer a projected regulatory motif (Fig. \ref{Info} right), which is derived in the Methods, is
\begin{equation*}
I_{\rm p}  = \textstyle \log_2 \left( \binom{N}{m} 2^{2^m} \right),
\end{equation*}
where $N$ is the number of genes in the network and $m = t_1 + \ldots + t_n$.
The information required to reverse engineer a regulatory motif (Fig. \ref{Info} left), also derived in the Methods, is
\begin{equation*}
I_{\rm r} = \textstyle \log_2 \left( \binom{N}{m} B_m \, c(t_1, \ldots, t_n) \right),
\end{equation*}
where $B_m$ is the $m$th Bell number.
\\ \indent
We show $I_{\rm r}$ and $I_{\rm p}$ for the 19 simplest regulatory motifs in Fig. \ref{Info}, for networks of 10 and 100 genes.
The mean of $I_{\rm r}$ is 24 bits and 42 bits for 10 genes and 100 genes, and
the mean of $I_{\rm p}$ is 80 and 98 bits for 10 and 100 genes.
This translates into a considerable savings in the experimental effort necessary to decode regulatory networks and parts thereof.
\\ \\ \noindent {\sf\textbf{\textcolor{red}{\large Methods}}} 
\\ \noindent{\sf\textbf{\textcolor{black}{Derivation of the number of biological logics}}}
\\ Here we derive an exact expression for the number of biologically permitted logics for different regulatory motifs.
This is the number of logics that can be expressed as a composition of logics, according to the dependence implied by each regulatory motif. 
\\ \indent
Let $q(t_1,\ldots,t_n)$ be the number of distinct compositions of logics that depend on at least one variable in each and every of the logics $g_i$ (Fig. \ref{TableCompositions}A).
The number of choices of $g_i$ that depend on at least one of its $t_i$ variables is $2^{2^{t_i}}-2$, since only true and false depend on no variables.
But, because of De Morgan duality, the set of logics $g_i$ and $\overline{g}_i$ are identical, so to avoid double counting we must divide this number by two.
Let 
\begin{equation*}
    \alpha_{t_i} = ( 2^{2^{t_i}}-2)/2.
\end{equation*}
Then
\begin{equation}
    q(t_1,\ldots,t_n)  =  s(n) \, \alpha_{t_1} \ldots \alpha_{t_n},
    \label{HA}  
\end{equation}
where $s(n)$ is defined in eq. (\ref{AllVariableLogics}).
For example,
\begin{eqnarray*}
    q(i)        	&=& 2 	\, \alpha_i,    						\\
    q(i,j)   		&=& 10 	\, \alpha_i \alpha_j,    				\\
    q(i,j,k)  	&=& 218 	\, \alpha_i \alpha_j \alpha_k.     
\end{eqnarray*}
We take $q(\emptyset)$ to be $s(0)$, which is 2.
\\ \indent 
To calculate the number of distinct logic compositions $c(t_1,\ldots,t_n)$, we just need to sum $q$ over the ways of depending on none of the $g_i$,
plus the ways of depending on just one of the $g_i$, and so on, up to the ways of depending on all $n$ of the $g_i$.
We can write this as
\begin{equation}
    c(t_1,\ldots,t_n)  =  	\!\!\!\!\!\! \sum_{e \in 2^{ \{ t_1,\ldots,t_n\} } } 	\!\!\!\!\!\! 	q(e),     
    \label{JA} 
\end{equation}
where the sum is over the power set of $\{t_1, \ldots, t_n\}$, that is, all subsets $e$ of the set $\{t_1, \ldots, t_n\}$, denoted by $2^{ \{ t_1,\ldots,t_n\} }$.
Inserting (\ref{HA}) into (\ref{JA}) gives
\begin{equation*}
    c(t_1,\ldots,t_n)  =  	\!\!\!\!\!\! \sum_{e \in 2^{ \{ t_1,\ldots,t_n\} }} 	\!\!\!\!\!\! 	s(|e|) \, \alpha_{\sigma_1} \ldots \alpha_{\sigma_{|e|}},    
\end{equation*}
where the $\sigma_i$ are the elements of $e$ and $|e|$ is the number of elements in $e$.
Grouping together subsets of the same size,
\begin{equation}
    c(t_1,\ldots,t_n)  =  \sum_{m=0}^n s(m)	\!\! \sum_{\sigma_1\ldots \sigma_m}   \, \alpha_{\sigma_1} \ldots \alpha_{\sigma_m},    
    \label{HI}
\end{equation}
as desired.
The second sum is over all subsets of size $m$ of the set $\{t_1, \ldots, t_n \}$. 
For $m=0$, the sum is over the null set and is taken to be 1.
\\ \\ \noindent{\sf\textbf{\textcolor{black}{Simulation of biological logics}}}
\\ To test our predictions, we simulated the logical dependence of one gene on other genes when they interact via transcription factors.
We wrote a program in Mathematica to handle each of the regulatory motifs in Fig. \ref{TableCompositions}B.
In particular, we determined the logical dependence of the top gene on the bottom genes for each possible assignment of logics to $f$ and to $g_1, \ldots, g_n$ (Fig. \ref{TableCompositions}A).
Since $f$ depends on $n$ variable, there are $2^{2^n}$ logics that $f$ must run through. 
Since $g_i$ depends on $t_i$ variables, there are $2^{2^{t_i}}$ logics that $g_i$ must run through, for each of the $g_i$. 
Thus we must run through a total of
\begin{equation*}
	2^{2^n} 2^{2^{t_1}}\!\!\ldots2^{2^{t_n}}
\end{equation*}
compositions of logics, for each of the regulatory motifs.
As an aside, this implies that $c(t_1,\ldots,t_n)$ is bounded from above by this number, which we indeed observed.
For example, for the regulatory motif $\{1,2,3\}$, we have $c(1,2,3)$ = 204,304 $\leq 2^{2^3} 2^{2^1} 2^{2^2} 2^{2^3} \!$ = 4,194,304.
\\ \\ \noindent{\sf\textbf{\textcolor{black}{Representation and composition of logics}}}
\\ Throughout this article, we write out logics in the disjunctive normal form, which consists of a disjunction of conjunctions.
In other words, we write them as {\sc or}s of {\sc and}s, or sums of products. 
As with the product and sum, {\sc and} takes precedence over {\sc or}.
Thus, for example, we have
\begin{center}
($a$ {\sc and} $b$) {\sc or} ($c$ {\sc and} $d$) $\equiv ab+cd$
\end{center}
and
\begin{center}
($a$ {\sc and} $c$) {\sc or} ($a$ {\sc and} $d$) {\sc or} ($b$ {\sc and} $c$) {\sc or} ($b$ {\sc and} $d$) \\
$\equiv ac + ad + bc + bd
\equiv (a+b)(c+d)$.
\end{center}

In general, for the regulatory motif $\{t_1,\ldots,t_n\}$, a logic is biologically permitted only if it can be expressed in the form
\begin{equation*}
\begin{aligned}
& h(x_{1,1},\ldots,x_{1,t_1}; \ldots; x_{n,1},\ldots,x_{n,t_n}) = \\
& f\big(g_1(x_{1,1},\ldots,x_{1,t_1}),\ldots,g_n(x_{n,1},\ldots,x_{n,t_n})\big),
\end{aligned}
\end{equation*}
where $x_{i,j}$ is the $j$th gene in the $i$th branch in Fig. \ref{TableCompositions}A.
For example, consider the bottom right logic in Fig. \ref{Valid}:
\begin{eqnarray*}
h(a,b,c,d) 	&=& (a b + \na \nb) (c d + \nc \nd) + (a \nb + \na b) (c \nd + \nc d) \\
	 	&=& (a b + \na \nb) (c d + \nc \nd) + \overline{(a b + \na \nb)} \,\, \overline{(c d + \nc \nd)} \\
		&=& g_1 g_2 + \overline{g_1} \,\, \overline{g_2}									
\end{eqnarray*}
where
\begin{equation*}
g_1(a,b) = a b + \na \nb \quad {\rm and} \quad g_2(c,d) = c d + \nc \nd.
\end{equation*}
Thus we are able to write $h(a,b,c,d)$ as a composition of a logic of two variables, each of which is a logic of two variables,
which corresponds to the regulatory motif $\{2,2\}$.
\\ \\ \noindent{\sf\textbf{\textcolor{black}{Information for reverse engineering}}}
\\ The information associated with realizing a discrete random variable that is uniformly distributed is $\log_2$ of the range of the random variable.
To reverse engineer a projected regulatory motif (Fig. \ref{Info} right), we need to deduce the connectivity and the logic, both of which we take to be uniformly distributed.
(Were the distributions to deviate from uniform, the required information would be less.)
For a network of $N$ genes, the number of ways that a gene can connect to $m = t_1 + \ldots + t_n$ other genes is $\binom{N}{m}$. 
The number of logics is $2^{2^m}$.
So the information, in bits, required to reverse engineer a projected regulatory motif is at most
\begin{equation*}
I_{\rm p}  = \textstyle \log_2 \left( \binom{N}{m} 2^{2^m} \right).
\end{equation*}
 \indent
Now let's reverse engineer a regulatory motif (Fig. \ref{Info} left).
For a network of $N$ genes, the number of ways that a gene can connect to other genes via transcription factors is $\binom{N}{m} B_m$, where $m = t_1 + \ldots + t_n$.
The second term is the $m$th Bell number, which is the number of ways to partition a set of $m$ labeled elements.
The number of logics is $c(t_1, \ldots, t_n)$ in eq. (\ref{CA}).
So the information required to reverse engineer a regulatory motif is at most
\begin{equation*}
I_{\rm r} = \textstyle \log_2 \left( \binom{N}{m} B_m \, c(t_1, \ldots, t_n) \right).
\end{equation*}
\\ 
\begin{scriptsize}%
\noindent{\sf\textbf{Acknowledgements}} \\
{\sf\textbf{Funding:}}
{\sf This research was supported by a grant from bit.bio.}
{\sf\textbf{Author contributions:}}
{\sf T. F. wrote the paper and did the mathematical derivations. T. F. and R. H. did the simulations and made the figures.}
{\sf\textbf{Competing interests:}}
{\sf The authors declare that they have no competing interests.}
{\sf\textbf{Acknowledgements:}}
{\sf The authors acknowledge Andriy Fedosyeyev and Alexander Mozeika for helpful discussions.}
\end{scriptsize}

\end{small}
\end{document}